\documentclass[a4paper,11pt]{article}
\usepackage[latin5]{inputenc}
\usepackage{hyperref}
\usepackage[onehalfspacing]{setspace}
\usepackage[top= 2cm, bottom= 2cm,a4paper]{geometry}
\usepackage{latexsym}
\usepackage{amsmath}
\usepackage{amsfonts}
\usepackage{amssymb}
\usepackage{color}
\usepackage{sectsty}
\usepackage{tabularx}
\usepackage{ragged2e}
\usepackage[labelfont=bf, labelsep=period]{caption}
\usepackage[sort&compress,numbers]{natbib}

\sectionfont{\fontsize{12}{15}\selectfont}
\subsectionfont{\fontsize{12}{15}\selectfont}

\def\NSCM/(A,B,\alpha){\mathsf{NSCM/(A,B,\alpha)}}
\def\NSGGd/G{\mathsf{NSGGd/G}}

\def\epsilon{\varepsilon}
\input{tcilatex}
\begin{document}

\title{\textbf{Application of Asymptotic Iteration Method (AIM) to a
Deformed Well Problem}}
\author{Hakan Ciftci\thanks{%
hciftci@gazi.edu.tr}$^{\text{ \ a}}$ and H.F. Kisoglu\thanks{%
hasanfatihk@aksaray.edu.tr}$^{\text{ \ b}}$ \\
$^{a}$\textit{Gazi Üniversitesi, Fen Fakültesi, Fizik Bölümü, 06500
Teknikokullar Ankara, Türkiye}\\
$^{b}$\textit{Physics Department, Faculty of Science and Letters, Aksaray
University, Aksaray, Turkey}}
\maketitle

\begin{abstract}
We have used Asymptotic Iteration Method (AIM) for obtaining the eigenvalues
of the Schrödinger's equation for a deformed well problem representing
trigonometric functions. By solving the problem, we have found that the
Schrodinger's equation for the considered potential has quasi-exact
solutions. Additionally, we have also calculated the perturbation expansion
of energy eigenvalues and found very simple analytical expression of the
energy. Finally, we have considered more general cases and obtained energy
eigenvalues for arbitrary potential parameters.
\end{abstract}


\section{Introduction}

The solution method of the Schrödinger's equation, which is the main scope
of quantum mechanics, depends upon the configuration of the system held,
i.e., behavior of the potential. One of the advantages of the asymptotic
iteration method (AIM), introduced in \cite{orj,1147}, is that it can be
used to solve the Schrödinger's equation exactly \cite{turbiner,oozer,yahoo,
ondort, onbes, onalti} or approximetly for numerous potentials \cite%
{trigonometric,boztosun,oozer2,022101,champion, onyedi}.\newline
AIM can also be used in the framework of perturbation theory. In common
perturbation theory, eigenfunctions of unperturbed hamiltonian (i.e. exact
solvable part) is required to construct the perturbation expansion for
energy eigenvalues and eigenfunctions. Furthermore, it is big challenging
even for third or fourth-order of approximation in some cases. The advantage
of the perturbation method using AIM is that the coefficients in
perturbation expansion can be calculated directly without any need to
eigenfunctions of unperturbed hamiltonian \cite%
{barakat,pert,022101,champion,2338}.\newline
We use the AIM to find out both the quasi-exact and approximate (numeric)
solutions of the Schrödinger's equation for a deformed well problem
representing trigonometric functions in the present work. The AIM is
summarized in section II. We find out the eigenvalues of Schrödinger's
equation approximately and quasi-exactly in section III for a potential
defined by

\begin{equation}
V(x)=\rule{0pt}{12pt}\rule{0cm}{0.42cm}A\func{Cos}(x)+B^{2}{\func{Sin}}%
^{2}(x)+\frac{\gamma (\gamma +1)}{\func{Sin}^{2}(x)},\quad x\in (0,\pi )
\label{intpot}
\end{equation}%
\newline
where $\mathit{A}$ and $\mathit{B}$ are coefficients and $\gamma $ ranges
over $\left( 0,\infty \right) $. In section IV, we apply perturbation method
for this kind of potential , just after, summarizing the perturbation theory
in the framework of AIM.

\section{The Asymptotic Iteration Method (AIM)}

Although the asymptotic iteration method has been introduced comprehensively
in \cite{orj}, we outline this method in this section according to
organization of the paper. The AIM can be used to solve second-order
homogeneous linear differential equations of the form

\begin{equation}
y^{\prime \prime }(x)=\lambda _{0}(x)y^{\prime }(x)+s_{0}(x)y(x)
\label{aimform}
\end{equation}%
\newline
where $\lambda _{0}$ and \textit{s}$_{0}$ have continuous derivatives in the
defined interval of the $\mathit{x}$ independent variable. According to
asymptotic aspect of the method, if we have

\begin{equation}
\frac{s_{n}}{s_{n-1}}=\frac{\lambda _{n}}{\lambda _{n-1}}\equiv \alpha
\label{asp}
\end{equation}%
\newline
for sufficient large $\mathit{n}>0$. The general solution of Eq.\ref{aimform}
is obtained as

\begin{equation}
y(x)=\exp \left( -\int\limits^{x}\alpha (t)dt\right) \left[
C_{2}+C_{1}\int\limits^{x}\exp \left( \int\limits^{t}\left( \lambda
_{0}(t)+2\alpha (t)\right) d\tau \right) dt\right]  \label{aimsol}
\end{equation}%
\newline
where

\QTP{Body Math}
\begin{eqnarray}
\lambda _{n} &=&\lambda _{n-1}^{\prime }+s_{n-1}+\lambda _{0}\lambda _{n-1}
\label{lames} \\
s_{n} &=&s_{n-1}^{\prime }+s_{0}\lambda _{n-1}  \notag
\end{eqnarray}%
\newline
Eigenvalues, \textit{E}, of the Schrödinger's equation are obtained using
the following termination condition by means of Eq.\ref{asp}

\begin{equation}
\delta _{n}(x)=s_{n}(x)\lambda _{n-1}(x)-\lambda _{n}(x)s_{n-1}(x)=0
\label{delta}
\end{equation}%
\newline
where the eigenvalues of \textit{n} energy levels are obtained by Eq.\ref%
{delta} with \textit{n} iterations, if the problem is analytically solvable
(exact solution). In this case, eigenvalues and eigenfunctions can be
determined in explicit algebraic form in AIM. Furthermore, there are limited
number of suitable potentials for this case.\newline
In exact solvable problem, condition $\delta _{n}=0$ is satisfied at every $%
\mathit{x}$ point\textit{\ }in the defined interval since $\delta _{n}$ is
independent from $\mathit{x}$. The (unknown) $\mathit{E}$ values are
determined from the $\mathit{n}$ roots of condition $\delta _{n}=0$. In case
of approximation, $\delta _{n}$ depends on both $\mathit{x}$ and $\mathit{E}$%
\textit{. }So, we have to determine a $\mathit{x}=\mathit{x}_{0}$ initial
value for solving $\delta _{n}=0$ with respect to \textit{E}. As this
initial $\mathit{x}_{0}$ value may be found from the minimum value of the
potential may also bounded domain of the eigenvalue problem \cite%
{trigonometric,champion}.

\section{AIM for the Problem}

We have considered a deformed well problem representing trigonometric
functions as below

\begin{equation*}
V(x)=\rule{0pt}{12pt}\rule{0cm}{0.42cm}A\func{Cos}(x)+B^{2}{\func{Sin}}%
^{2}(x)+\frac{\gamma (\gamma +1)}{\func{Sin}^{2}{x}},\quad x\in (0,\pi )
\end{equation*}%
\newline
where $\mathit{A}$, $\mathit{B}$ and $\gamma $ are the potential parameters. 
$\gamma $ is defined in the interval $\left[ 0,\infty \right) .$ If we
substitute $\mathit{V}(x)$ in the Schrödinger's equation, we have

\begin{equation}
\left\{ \frac{d^{2}}{dx^{2}}+E-\left[ A\func{Cos}(x)+B^{2}{\func{Sin}}%
^{2}(x)+\frac{\gamma (\gamma +1)}{\func{Sin}^{2}(x)}\right] \right\} \Psi
(x)=0  \label{potsch}
\end{equation}%
\newline
where unnormalized eigenfunction $\Psi (x)\in $ $\mathit{L}^{2}\left( 0,\pi
\right) $. We may assume that $\Psi (x)$ takes the form\newline
\begin{equation}
\Psi (x)={\func{Sin}}^{\gamma +1}\left( x\right) e^{B\func{Cos}\left(
x\right) }f\left( x\right)  \label{potsol}
\end{equation}%
\newline
with this choice, $\Psi (x)$ vanishes at $\mathit{x}=0$ and $\mathit{x}=\pi $%
. Once we substitute above equation in Eq.\ref{potsch}\newline
\begin{equation}
f^{\prime \prime }(x)+2\left[ \left( \gamma +1\right) \frac{\func{Cos}(x)}{%
\func{Sin}(x)}-B\func{Sin}(x)\right] f^{\prime }(x)+\left[ E-\left( \gamma
+1\right) ^{2}-\left\{ A+B\left( 2\gamma +3\right) \right\} \func{Cos}(x)%
\right] f(x)=0  \label{solab}
\end{equation}%
\newline
is obtained. If we change the variable such as $\func{Cos}(x)=y$ in Eq.\ref%
{solab} which yields

\begin{equation}
f^{\prime \prime }(y)=2\left( \frac{\sigma y}{1-y^{2}}-B\right) f^{\prime
}(y)+\left( \frac{\xi y-\omega }{1-y^{2}}\right) f(y)  \label{potaimon}
\end{equation}%
\newline
where $\omega =E-\left( \gamma +1\right) ^{2}$, $\xi =A+2\sigma B$ and $%
2\sigma =2\gamma +3$. The last equation is in form of Eq.\ref{aimform} for
which AIM can be used. We obtain the (unkown) \textit{E}\emph{\ }energy
eigenvalues by using AIM.

\subsection{Quasi-Exact Solutions of the Problem}

We have obtained Eq.\ref{potaimon} for using the AIM for determining the 
\textit{E} eigenvalues of Eq.\ref{potsch}. Our calculations show that the
problem has quasi-exact solution under some special conditions on the
potential parameters. AIM gives us that if $\xi =-2nB$ with $n=0,1,2,3,...$
the energy eigenvalues can be obtained exactly. We will give first two
results analytically because of the complecity of the calculations, but we
will give more numerical results in Table.\ref{er}. Using $\xi =-2nB$ and $%
\xi =A+2\sigma B$ together, we get the following formula

\begin{equation}
A=-2B\left( n+\gamma +\frac{5}{2}\right)  \label{link}
\end{equation}%
\newline
For $n=0$, we get $\omega =0$ and $E_{00}=(\gamma +1)^{2}$. For $n=1$, we
get $\omega =\sigma \mp \sqrt{\sigma ^{2}+4B^{2}}$ and the energies can be
obtained as follows

\begin{eqnarray}
E_{10} &=&\left( \gamma +\frac{3}{2}\right) ^{2}+\frac{1}{4}-\sqrt{\left(
\gamma +\frac{3}{2}\right) ^{2}+4B^{2}} \\
E_{11} &=&\left( \gamma +\frac{3}{2}\right) ^{2}+\frac{1}{4}+\sqrt{\left(
\gamma +\frac{3}{2}\right) ^{2}+4B^{2}}
\end{eqnarray}
Numerically calculated eigenvalues for $\mathit{B}=1$ and $\sigma =4.5$ can
be found in Table.\ref{er}.

\bigskip

\begin{table}[h]
\caption{Numerically calculated eigenvalues for $B=1$ and $\protect\sigma %
=4.5$ for $n=0$ to 5 by using\emph{\ }$\protect\xi =-2nB$. $k$ is sub-level
for each $n$ level and $k$=0, 1, 2, ..., $n$.}
\label{er}\centering%
\begin{tabular}{|c|c|c|c|c|}
\hline
$n$ & $k$ & $A$ & $\omega _{nk}$ & $E_{nk}$ \\ \hline
0 & 0 & -11 & 0 & 16.0000 \\ \hline
1 & 0 & -13 & -0.424429 & 15.5756 \\ 
& 1 & -13 & 9.42443 & 25.4244 \\ \hline
& 0 & -15 & -0.919071 & 15.0809 \\ 
2 & 1 & -15 & 9.27711 & 25.2771 \\ 
& 2 & -15 & 20.642 & 36.642 \\ \hline
& 0 & -17 & -1.47815 & 14.5218 \\ 
3 & 1 & -17 & 9.09552 & 25.0955 \\ 
& 2 & -17 & 20.6205 & 36.6205 \\ 
& 3 & -17 & 33.7621 & 49.7621 \\ \hline
& 0 & -19 & -2.09614 & 13.9039 \\ 
& 1 & -19 & 8.87768 & 24.8777 \\ 
4 & 2 & -19 & 20.5892 & 36.5892 \\ 
& 3 & -19 & 33.7953 & 49.7953 \\ 
& 4 & -19 & 48.834 & 64.834 \\ \hline
& 0 & -21 & -2.76786 & 13.2321 \\ 
& 1 & -21 & 8.62229 & 24.6223 \\ 
& 2 & -21 & 20.5459 & 36.5459 \\ 
5 & 3 & -21 & 33.8293 & 49.8293 \\ 
& 4 & -21 & 48.8905 & 64.8905 \\ 
& 5 & -21 & 65.8799 & 81.8799 \\ \hline
\end{tabular}%
\end{table}

\subsection{Approximately Determination of the Eigenvalues}

\label{3.2}

In this case, we have determined the eigenvalues for arbitrary $\mathit{A}$, 
$\mathit{B}$ and $\gamma $ values. Calculated eigenvalues ($\mathit{E}%
_{direct}$) of $\mathit{n}=0$ to $5$ are given in Table.\ref{t2} in which $%
\mathit{A}$, $\mathit{B}$ and $\gamma $ (in ($\mathit{A}$; $\mathit{B}$; $%
\gamma $) form) are ($0.167;0.0019;0.0019$), ($2;0.5;$1) and ($4;1;1.97$).
In the calculation we take $y_{0}=\frac{1}{2}$.

\bigskip

\begin{table}[h]
\caption{Approximately calculated eigenvalues $E_{direct}$ for (0.167;
0.0019; 0.0019), (2; 0.5; 1) and (4; 1; 1.97) in ($A$; $B$; $\protect\gamma $%
) form. The value written as subscript for each eigenvalue represents the
iteration number.}
\label{t2}\centering%
\begin{tabular}{|c|c|c|c|c|c|c|}
\hline
& \multicolumn{2}{|c|}{$A=0.167$, $B=\gamma =0.0019$} & \multicolumn{2}{|c|}{%
$A=2$, $B=0.5$, $\gamma =1$} & \multicolumn{2}{|c|}{$A=4$, $B=1$, $\gamma
=1.97$} \\ \hline
$n$ & $\omega $ & $E_{direct}$ & $\omega $ & $E_{direct}$ & $\omega $ & $%
E_{direct}$ \\ \hline
0 & -0.00231 & $1.00149_{(3)}$ & 0.07399 & $4.07399_{(10)}$ & 0.57328 & $%
9.39418_{(13)}$ \\ \hline
1 & 3.00473 & $4.00853_{(4)}$ & 5.17104 & $9.17104_{(12)}$ & 7.618 & $%
16.4389_{(13)}$ \\ \hline
2 & 8.008 & $9.0118_{(5)}$ & 12.1609 & $16.1609_{(11)}$ & 16.5264 & $%
25.3473_{(14)}$ \\ \hline
3 & 15.0116 & $16.0154_{(6)}$ & 21.1506 & $25.1506_{(13)}$ & 27.4331 & $%
36.254_{(16)}$ \\ \hline
4 & 24.0153 & $25.0191_{(7)}$ & 32.1438 & $36.1438_{(14)}$ & 40.348 & $%
49.1689_{(18)}$ \\ \hline
5 & 35.0191 & $36.0229_{(8)}$ & 45.1392 & $49.1392_{(15)}$ & 55.2698 & $%
64.0907_{(19)}$ \\ \hline
\end{tabular}%
\end{table}

\bigskip

\noindent As is seen from Table.\ref{t2}, energy eigenvalues act as $\sim
(n+\gamma +1)^{2}$.

\section{Perturbation Theory within the frame of AIM}

Although the usage of perturbation method within AIM is introduced in \cite%
{pert} comprehensively, we give a summary about the application method in
this section. Assume that the potential of a system is written in a form of

\begin{equation}
V(x)=V_{0}(x)+\mu V_{p}(x)  \label{pertpot}
\end{equation}%
\newline
where $V_{0}(x)$ is solvable (or unperturbed hamiltonian) potential. $%
V_{p}(x)$ and $\mu $ are potential of the perturbed hamiltonian and
perturbation expansion parameter, respectively. So, the Schrödinger's
equation reads

\begin{equation}
\left( -\frac{d^{2}}{dx^{2}}+V_{0}(x)+\mu V_{p}(x)\right) \Psi (x)=E\Psi (x)
\label{pertsch}
\end{equation}%
\newline
where $E_{n}$ eigenvalues are written as the expansion of $j$ th-order
correction $E_{n}^{(j)}$ as follows

\begin{equation}
E_{n}=E_{n}^{(0)}+\mu E_{n}^{(1)}+\mu
^{2}E_{n}^{(2)}+...=\sum\limits_{j=0}^{\infty }\mu ^{j}E_{n}^{(j)}
\label{perteig}
\end{equation}%
\newline
The form of Eq.\ref{aimform} can be yielded by a substitution of (in general
form) $\psi (x)=\psi _{0}(x)f(x)$ in Eq.\ref{pertsch}. So, one can obtain
the following equation for $f(x)$

\begin{equation}
f^{\prime \prime }(x)=\lambda _{0}(x,\mu ,E)f^{\prime }(x)+s_{0}(x,\mu
,E)f(x)  \label{pertAIM}
\end{equation}%
\newline
The termination condition, Eq.\ref{delta}, is written as

\begin{equation}
\delta _{n}(x,\mu ,E)=s_{n}(x,\mu ,E)\lambda _{n-1}(x,\mu ,E)-\lambda
_{n}(x,\mu ,E)s_{n-1}(x,\mu ,E)=0  \label{pertdelta}
\end{equation}%
\newline
Once $\delta _{n}(x,\mu ,E)$ is expanded about $\mu =0$, we obtain

\begin{equation}
\delta _{n}(x,\mu ,E)=\delta _{n}(x,0,E)+\left. \frac{\mu }{1!}\frac{%
\partial \delta _{n}(x,\mu ,E)}{\partial \mu }\right\vert _{\mu =0}+\left. 
\frac{\mu ^{2}}{2!}\frac{\partial ^{2}\delta _{n}(x,\mu ,E)}{\partial \mu
^{2}}\right\vert _{\mu =0}+...=\sum\limits_{k=0}^{\infty }\mu ^{k}\delta
_{n}^{(k)}(x,E)=0  \label{pertdeltaseri}
\end{equation}%
\newline
where $\delta _{n}^{(k)}(x,E)=\left. \frac{1}{k!}\frac{\partial ^{k}\delta
_{n}(x,\mu ,E)}{\partial \mu ^{k}}\right\vert _{\mu =0}$. The condition of

\begin{equation}
\delta _{n}^{(k)}(x,E)=0  \label{deltak}
\end{equation}%
\newline
can be obtained by means of Eq.\ref{pertdeltaseri} for each $k$. According
to the perturbation method in the framework of AIM, solving the equation $%
\delta _{n}(x,0,E)=0$ with respect to (unknown) $E$ gives $E_{n}^{(0)}$
(eigenvalues of unperturbed hamiltonian), equation $\delta _{n}^{(1)}(x,E)=0$
gives $E_{n}^{(1)}$ (first-order correction to $E_{n}$), $\delta
_{n}^{(2)}(x,E)$ gives $E_{n}^{(2)}$ (second-order correction to $E_{n}$)
and so on. The eigenfunctions can also be found with similar manner. This is
an attractive feature of the AIM usage in the perturbation theory for
obtaining the eigenfunctions $f_{n}(x)$ given as follows

\begin{equation}
f_{n}(x)=\exp \left( -\int\limits^{x}\alpha _{n}(t,\mu )dt\right)
\label{perteigenfunc}
\end{equation}%
\newline
where $\alpha _{n}(t,\mu )\equiv s_{n}(t,\mu )/\lambda _{n}(t,\mu )$. $%
\alpha _{n}(t,\mu )$ is expanded about $\mu =0$ in a similar manner of
obtaining the eigenvalues. So,

\begin{equation}
\alpha _{n}(t,\mu )=\sum\limits_{k=0}^{\infty }\mu ^{k}\alpha _{n}^{(k)}(t)
\label{pertalfa}
\end{equation}%
\newline
where\ $\alpha _{n(x)}^{(k)}=\left. \frac{1}{k!}\frac{\partial ^{k}\alpha
_{n(x,\mu )}}{\partial \mu ^{k}}\right\vert _{\mu =0}$. Thus, perturbation
expansion of the $f_{n(x)}$ is written as follows

\begin{equation}
f_{n}(x)=\exp \left[ \sum\limits_{k=0}^{\infty }\mu ^{k}\left(
-\int\limits^{x}\alpha _{n}^{(k)}(t)dt\right) \right] =\prod\limits_{k=0}^{%
\infty }f_{n}^{(k)}(x)  \label{pertexpeigenfunction}
\end{equation}%
\newline
where $k$ th-order correction $f_{n}^{(k)}(x)$ to $f_{n}(x)$ is

\begin{equation}
f_{n}^{(k)}(x)=\mu ^{k}\left( -\int\limits^{x}\alpha _{n}^{(k)}(t)dt\right)
\label{pertcorrectoeigenfunc}
\end{equation}

\subsection{Perturbation Theory for the Potential Function in the frame of
AIM}

In section III, we have obtained the eigenvalue equation in the appropriate
form for AIM usage as follows

\begin{equation}
f^{\prime \prime }(y)=2\left( \frac{\sigma y}{1-y^{2}}-B\right) f^{\prime
}(y)+\left( \frac{\xi y-\omega }{1-y^{2}}\right) f(y)  \label{potaim}
\end{equation}%
\newline
where $\omega =E-\left( \gamma +1\right) ^{2}$, $\xi =A+2\sigma B$ and $%
2\sigma =2\gamma +3$. $A$ and $B$ parameters of the potential $V(x)$, given
by Eq.\ref{intpot}, was linked to eachother via Eq.\ref{link}. Suppose that $%
A=2aB$, in this case $\xi =2(a+\sigma )B$ and the differential equation can
be written as below

\begin{equation}
f^{\prime \prime }(y)=2\left( \frac{\sigma y}{1-y^{2}}-B\right) f^{\prime
}(y)+\left( \frac{2(a+\sigma )By-\omega }{1-y^{2}}\right) f(y)
\end{equation}
\newline
Now, $B$ can be taken as perturbation expansion parameter. We can expand $%
\omega $ as below

\begin{equation}
\omega _{n}=\omega _{n}^{(0)}+B\omega _{n}^{(1)}+B^{2}\omega _{n}^{(2)}+...
\label{omegapert}
\end{equation}%
\newline
In this expansion the general form of zeroth-order correction $\omega
_{n}^{(0)}$ is obtained via,

\begin{equation}
\delta ^{(0)}(x,0,\omega _{n}^{(0)})=0  \label{deltazerothorder}
\end{equation}%
\newline
as mentioned previously. The obtained few values of $\omega _{n}^{(0)}$ are $%
0$ , $2\sigma $, $2+4\sigma $, $6+6\sigma $, $12+8\sigma $, $20+10\sigma $,
respectively. One can easily write general formula for $\omega _{n}^{(0)}$
as below

\begin{equation}
\omega _{n}^{(0)}=n\left( n-1\right) +2n\sigma  \label{omega0}
\end{equation}

\bigskip

\noindent The first-order correction to $\omega _{n}$ is generalized in the
similar manner, and it is found that $\omega _{n}^{(1)}=0$ for each $n$
level via the equation $\delta ^{(1)}(x,0,\omega _{n}^{(1)})=0$. The
second-order correction to $\omega _{n}$ is generalized as

\begin{equation}
\omega _{n}^{(2)}=-2\left( \frac{a^{2}g_{1}(n,\sigma )-g_{2}(n,\sigma )}{%
g_{3}(n,\sigma )}\right)  \label{omega2}
\end{equation}%
\newline
where

\begin{eqnarray}
g_{1}(n,\sigma ) &=&2\sigma ^{2}-\left( 2n+5\right) \sigma -\left(
n^{2}-n-3\right)  \notag  \label{g1g2g3} \\
g_{2}(n,\sigma ) &=&2\sigma ^{4}+\left( 6n-5\right) \sigma ^{3}+\left(
7n^{2}-11n+3\right) \sigma ^{2}+4n\left( n-1\right) ^{2}\sigma +n^{2}\left(
n-1\right) ^{2} \\
g_{3}(n,\sigma ) &=&\left( 2\sigma +2n+1\right) \left( 2\sigma +2n-3\right)
\left( \sigma +n\right) \left( \sigma +n-1\right)  \notag
\end{eqnarray}%
\newline
If Eq.\ref{omega0} and Eq.\ref{omega2} are substituted in Eq.\ref{omegapert}
one can obtain

\begin{equation}
\omega _{n}=n\left( n-1\right) +2n\sigma -\left( \frac{A^{2}g_{1}(n,\sigma
)-4B^{2}g_{2}(n,\sigma )}{2g_{3}(n,\sigma )}\right)  \label{omegason}
\end{equation}%
\newline
where we use $a=\frac{A}{2B}$ substituted in Eq.\ref{omega2} as well. From $%
\omega _{n}=E_{n}-\left( \gamma +1\right) ^{2}$, we get $E_{n}^{p}$
(perturbed energy) approximately as follows

\begin{equation}
E_{n}^{p}\approx \left( \gamma +n+1\right) ^{2}-\left( \frac{%
A^{2}g_{1}(n,\sigma )-4B^{2}g_{2}(n,\sigma )}{2g_{3}(n,\sigma )}\right)
\label{pertenerji}
\end{equation}%
\newline
Numeric results are given in Table.\ref{t3} in which $E_{n}^{p}$, given in
Eq.\ref{pertenerji}, was used for $n=0$ to $5$, and in Table.\ref{t4} for $%
n=0$ to $10$. The comparison of $E_{n}^{p}$ with numerically calculated
eigenvalues $E_{direct}$ for arbitrary $A$, $B$ and $\gamma$ is also given
in Table.\ref{t3} and Table.\ref{t4}. The values written as subscript
represent the iteration numbers in both tables.

\bigskip

\begin{table}[h]
\caption{Comparison of eigenvalues of perturbed hamiltonian $E_{n}^{p}$ with
numerically ones $E_{direct}$ for $A=2$, $B=2$ and $\protect\gamma =4$ for $%
n=0$ to $5$.}
\label{t3}\centering%
\begin{tabular}{|c|c|c|c|}
\hline
$n$ & $\omega _{n}$ & $E_{direct}$ & $E_{n}^{p}$ \\ \hline
0 & 3.6259 & $28.6259_{(13)}$ & 28.6364 \\ \hline
1 & 14.1219 & $39.1219_{(14)}$ & 39.1329 \\ \hline
2 & 26.8245 & $51.8245_{(17)}$ & 51.8308 \\ \hline
3 & 41.6323 & $66.6323_{(18)}$ & 66.6353 \\ \hline
4 & 58.5005 & $83.5005_{(19)}$ & 83.5015 \\ \hline
5 & 77.406 & $102.406_{(19)}$ & 102.406 \\ \hline
\end{tabular}%
\end{table}

\bigskip

\begin{table}[h]
\caption{Comparison of the eigenvalues of perturbed hamiltonian $E_{n}^{p}$
with numerically ones $E_{direct}$ for $A=2$, $B=1$ and $\protect\gamma %
=0.0019$ for $n=0$ to $10$.}
\label{t4}\centering%
\begin{tabular}{|c|c|c|c|}
\hline
$n$ & $\omega _{n}$ & $E_{direct}$ & $E_{n}^{p}$ \\ \hline
0 & 0.38637 & $1.39017_{(12)}$ & 1.42145 \\ \hline
1 & 3.65921 & $4.66301_{(13)}$ & 4.64073 \\ \hline
2 & 8.56603 & $9.56983_{(16)}$ & 9.56855 \\ \hline
3 & 15.5445 & $16.5483_{(15)}$ & 16.547 \\ \hline
4 & 24.5364 & $25.5402_{(16)}$ & 25.5392 \\ \hline
5 & 35.5338 & $36.5376_{(18)}$ & 36.5368 \\ \hline
6 & 48.5336 & $49.5374_{(18)}$ & 49.5369 \\ \hline
7 & 63.5349 & $64.5387_{(19)}$ & 64.5383 \\ \hline
8 & 80.537 & $81.5408_{(20)}$ & 81.5404 \\ \hline
9 & 99.5392 & $100.543_{(20)}$ & 100.543 \\ \hline
10 & 120.542 & $121.546_{(21)}$ & 121.546 \\ \hline
\end{tabular}%
\end{table}

\bigskip

\noindent It can be seen from Table.\ref{t3} that numerically calculated
eigenvalues $E_{direct}$ are consistent with the ones of perturbed
hamiltonian $E_{n}^{p}$. Also, they are more consistent as the energy level $%
n$ increases, and the values of $E_{direct}$ converge more slowly with
increasing $n$. All these comments apply to Table.\ref{t4} as well.
Futhermore, the eigenvalues act as $\sim (n+\gamma +1)^{2}$ while $\omega
_{n}\approx n(n+2)$ in Table.\ref{t4}.\newline
In Table.\ref{t6}, values of $E_{direct}$ and $E_{n}^{p}$ are inconsonant by
selecting $A=B=5$ and $\gamma =0.5$.

\bigskip

\begin{table}[h]
\caption{Comparison of the eigenvalues of perturbed hamiltonian $E_{n}^{p}$
with numerically calculated ones $E_{direct}$ for $A=B=5$ and $\protect%
\gamma =0.5$ for $n=0$ to $5$. The values written as subscript represent the
iteration numbers.}
\label{t6}\centering%
\begin{tabular}{|c|c|c|c|}
\hline
$n$ & $\omega _{n}$ & $E_{direct}$ & $E_{n}^{p}$ \\ \hline
0 & 11.8935 & $14.1435_{(31)}$ & 21.0000 \\ \hline
1 & 19.3048 & $21.5548_{(34)}$ & 20.8333 \\ \hline
2 & 25.0743 & $27.3243_{(36)}$ & 25.7917 \\ \hline
3 & 32.0111 & $34.2611_{(40)}$ & 33.375 \\ \hline
4 & 41.5015 & $43.7515_{(37)}$ & 43.1667 \\ \hline
5 & 53.2222 & $55.4722_{(39)}$ & 55.0476 \\ \hline
\end{tabular}%
\end{table}

\section{Conclusion}

We have used AIM for solving Schrödinger's equation for a deformed well
problem representing trigonometric function. In this study, we have applied
AIM to quasi-exact solvable eigenvalue problem for obtaining the energy
eigenvalues and seen that each energy level $n$ has a sub-level $k$ that $%
k=0,1,2,...,$ $n$. Method has also been used to obtain the eigenvalues
approximately by using arbitrary potential parameters. According to results,
we can say that the energy eigenvalues behave as $\sim (n+\gamma +1)^{2}$
for larger quantum number $n$, as is seen in Eq.\ref{pertenerji}. AIM can
also be used in the framework of perturbation theory. In common perturbation
theory, eigenfunctions of unperturbed hamiltonian is required to construct
the perturbation expansion for energy eigenvalues and eigenfunctions. Also,
determining the third or fourth-order correction to energy eigenvalues is
usually quite big challenging. The advantage of the perturbation method
using AIM is that the coefficients in perturbation expansion can be
calculated directly without any need to eigenfunctions of unperturbed
hamiltonian and any complication.\newline
Perturbation method has also been applied to potential function held in
scope of the AIM, in this study. According to results, eigenvalues of
perturbed hamiltonian are consistent with those of approximately calculated
ones for which arbitrary parameters of the potential have been used.
However, they are inconsonant for some set of the parameters. Although this
inconsistency is true for $n=0$ to $5$, it tends to decrease as $n$
increases.\newline
Consequently, it can be said that we can make the eigenvalues tend to
specific behavior by selecting appropriate potential parameters.

\end{document}